\documentclass[twocolumn]{webofc}

\usepackage[varg]{txfonts}   
\usepackage{subfigure}
\usepackage{amsmath}
\usepackage{bm}
\usepackage[T1]{fontenc}
\usepackage{graphicx}
\usepackage{color}

\begin{document}
\title{Bulk modulus of soft particle assemblies under compression}

\author{\firstname{David} \lastname{Cantor}\inst{1}\fnsep\thanks{\email{david.cantor@polymtl.ca}} 
        \and
        \firstname{Manuel} \lastname{C\'{a}rdenas-Barrantes}\inst{2}
        \and
        \firstname{Itthichai} \lastname{Preechawuttipong}\inst{3}
        \and
        \firstname{Mathieu} \lastname{Renouf}\inst{2}
        \and
        \firstname{Emilien} \lastname{Az\'{e}ma}\inst{2,4}
}

\institute{Department of Civil, Geological and Mining Engineering, Polytechnique Montr\'{e}al, Qu\'{e}bec, Canada
\and
LMGC, Universit\'{e} de Montpellier, CNRS, Montpellier, France
\and
Department of Mechanical Engineering, Chiang Mai University, Chiang Mai, Thailand
\and
Institut Universitaire de France (IUF), Paris, France
}

\abstract{
Using a numerical approach based on the coupling of the discrete and finite element methods, we explore the variation of the bulk modulus $K$ of soft particle assemblies undergoing isotropic compression. 
As the assemblies densify under pressure-controlled boundary conditions, we show that the non-linearities of $K$ rapidly deviate from predictions standing on a small-strain framework or the, so-called, Equivalent Medium Theory (EMT). 
Using the granular stress tensor and extracting the bulk properties of single representative grains under compression, we propose a model to predict the evolution of $K$ as a function of the sample's solid fraction and a reference state as the applied pressure $P \rightarrow 0$. 
The model closely reproduces the trends observed in our numerical experiments confirming the behavior scalability of soft particle assemblies from the individual particle scale. 
Finally, we present the effect of the interparticle friction on $K$'s evolution and how our model easily adapts to such a mechanical constraint. 
}
\maketitle
\section{Introduction}\label{intro}
Soft granular particulate materials such as powders, gels, bubbles, rubber chunks, and even cells are challenging materials to characterize and model due to the large deformation they can undergo. 
Besides, their discrete nature calls for adequate interaction laws for multi-contact systems. 
Amongst the different numerical approaches to model highly deformable particle assemblies, the discrete-element approach is one of the most frequently used. 
By means of smoothed interaction laws between bodies, simulations may reproduce some elasticity due to a virtual contact deflection, although the bodies themselves do not undergo strains. 
Using this approach, the elastic properties of particle assemblies have been studied, however restricting considerations of the small-strain domain of deformations \cite{Agnolin2007c,Gu2013,Khalili2017b,VanderWerf2020}. 
More recently, the development of more advanced methods coupling discrete and finite elements \cite{Vu2019,Guner2015,Procopio2005,Abdelmoula2017,Huang2017,Mollon2018} or meshless methods \cite{Boromand2018,Nezamabadi2019} have permitted to explore the compression behavior of soft granular media beyond jamming.
Nonetheless, the study of the evolution of elastic properties of particulate assemblies undergoing large deformation is still challenging to characterize. 

In this paper, we simulate assemblies of 2D circular particles under isotropic compression using the contact dynamics method and the finite-element method to account for the large deformation of meshed bodies. 
Systematically increasing the pressure on assemblies of disks, we have access to the macroscopic stress-strain relation letting us deduce the evolution of the bulk modulus during the compaction. 
Then, using the granular stress tensor and the behavior of individual representative particles, we propose an analytical equation for the bulk modulus evolution fitting very well our numerical experiments. 

In Sec. \ref{Numerics}, we present the details of the numerical approach and the isotropic test procedures. 
Then, in Sec. \ref{Macro}, we show the stress-strain relation measured at the sample scale and the bulk modulus evolution. 
In order to look for the origins of the macroscopic bulk modulus, we explore in Sec. \ref{Local} the behavior of individual particles under compression. 
In Sec. \ref{Model}, we introduce an analytical approach based on the stress tensor decomposition from microstructural parameters letting us deduce an equation for the macroscopic bulk behavior. 
Finally, in Sec. \ref{Friction}, we present the effect of the interparticle coefficient of friction on the evolution of the bulk modulus and how our analytical model easily adapts to such mechanical constraint.
Section \ref{Conclu} concludes this work with a summary and perspectives. 

\section{Numerical procedures}\label{Numerics}
To simulate assemblies of soft particles, we used the coupling of the discrete method known as contact dynamics (CD) \cite{Jean1992,Dubois2018} and classic finite elements in the framework coined as non-smooth contact dynamics (NSCD) by M. Jean \cite{Jean1999}. 
With this method, we were capable of building assemblies of circular meshed bodies that interact using unilateral contacts and dry friction. 

\begin{figure}
    \centering
    \subfigure {\label{fig:ex1}
    \includegraphics[width=0.47\linewidth]{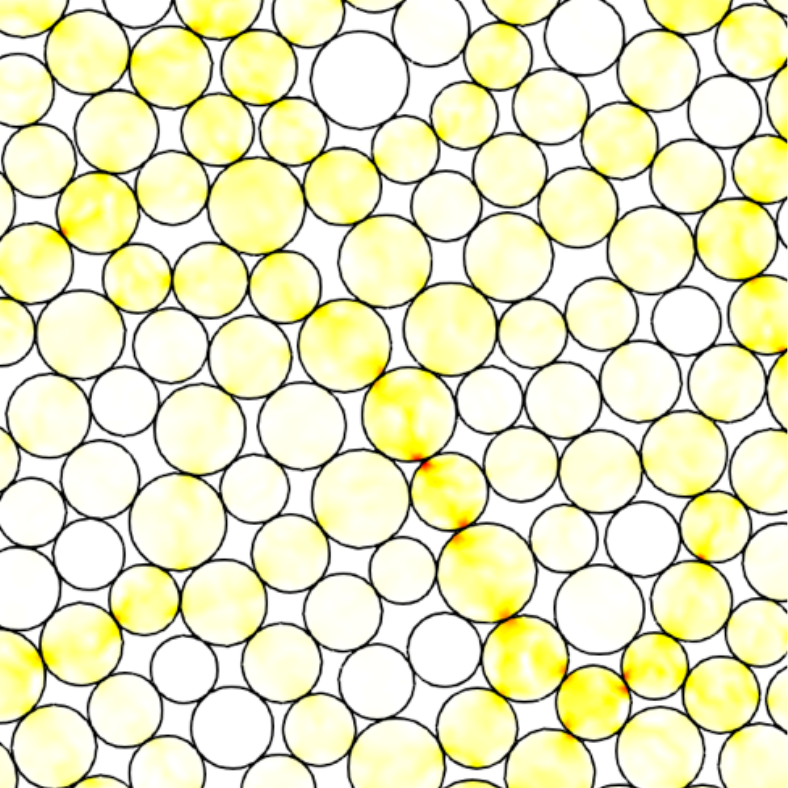}}
    \subfigure {\label{fig:ex2}
    \includegraphics[width=0.47\linewidth]{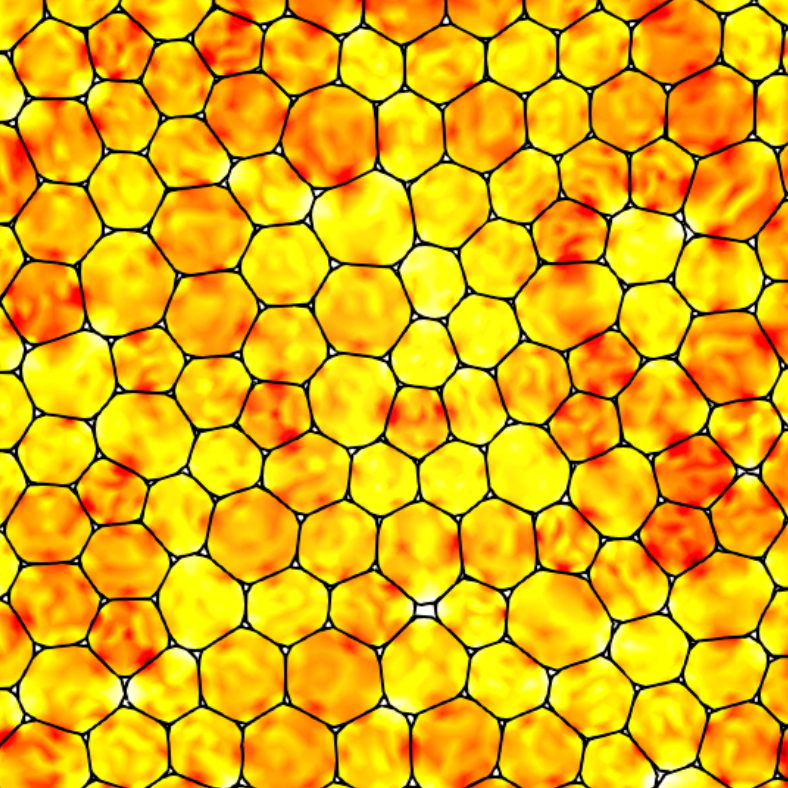}}
    \caption{Screenshots of a sample under a relative pressure $P/E = 5 \times 10^{-4}$ (left), and $P/E = 5 \times 10^{-1}$ (right). The intensity of the color is proportional to the volumetric strain within the particles.} 
    \label{fig:shots}
\end{figure}

We built samples composed of $N_p = 1500$ circular bodies slightly disperse in size (the ratio between the maximal over the minimal particle diameter is $1.5$), meshed using 92 triangular finite elements, and deposited within square boxes by means of an algorithm based on simple geometrical properties. 
For the finite elements, we used the non-Hookean hyper-elastic material model \cite{Rivlin1948}, setting incompressible bulk behavior, plane strain conditions, and an elastic modulus $E$. 
We previously tested the mesh resolution concluding that the number of finite elements per grain we ended up using does not compromise the results' quality. 
Then, we set pressure controlled conditions on the four rigid walls of the boxes in a series of steps varying the relative pressure $P/E$ in the range $[1\times 10^{-4}, 5 \times 10^{-1}]$.
The interparticle coefficient of friction $\mu$ was set to zero and gravity was neglected to avoid pressure gradients. 
Also, note that the loading was performed using slow gradual steps to avoid dynamic effects and promote a rapid dissipation of elastic waves. 
Figure \ref{fig:shots} shows the sample's configuration at the beginning of the loading and its deformed state at the end of the tests. 

\section{Macroscopic bulk modulus}\label{Macro}
In our simulations, we can easily track the deformations of the servo-controlled boundary walls. 
For convenience, we relate the volumetric strain of the sample $\varepsilon_v$ to the solid fraction $\phi$, as $\varepsilon_v = -\ln(\phi_0/\phi)$, with $\phi_0$ being the solid fraction as the ratio $P/E \rightarrow 0$. 
This reference state can be understood either as the solid fraction the assembly presents with perfectly rigid bodies or with vanishing external pressure. 
The macroscopic bulk modulus $K$ can be then computed as a function of the solid fraction as $K = (dP/d\phi)(d\phi/d\varepsilon_v)$. 

\begin{figure}
    \centering
    \includegraphics[width=0.9\linewidth]{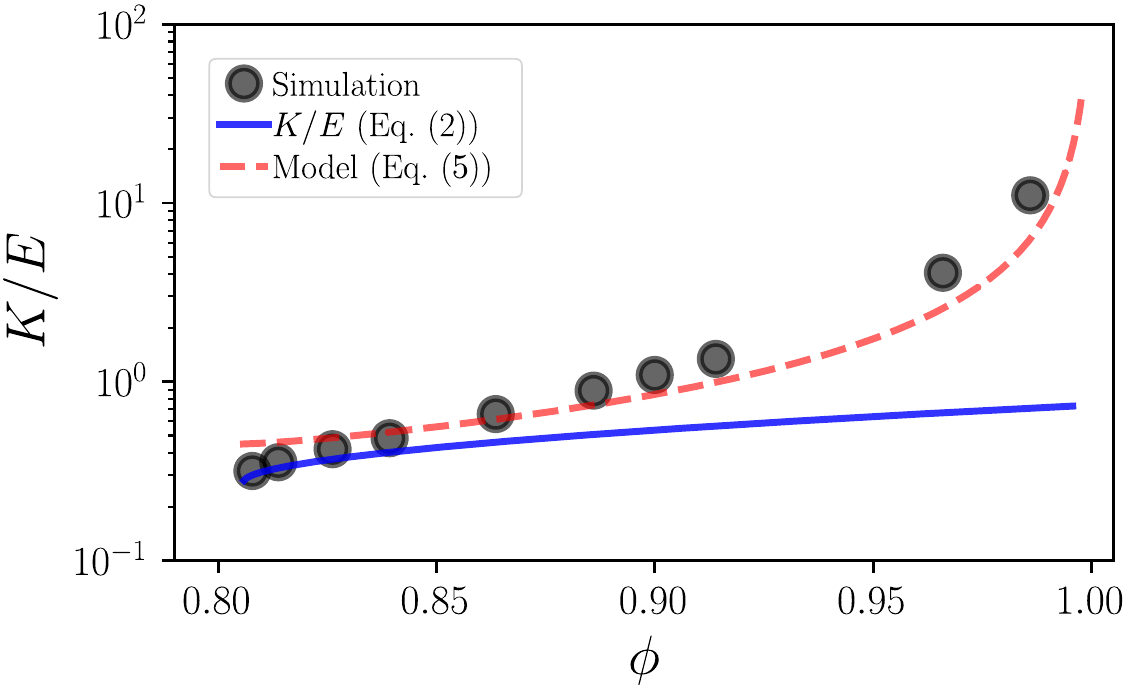}
    \caption{Evolution of the bulk modulus during compaction in our experiments (black disks), and the predictions given by Eqs. (\ref{eq:k1}) and (\ref{eq:k2}) with a dashed blue and red lines, respectively.} 
    \label{fig:k_and_eqs}
\end{figure}

Figure \ref{fig:k_and_eqs} presents the evolution of $K$ as a function of $\phi$. 
Different strategies have been used to predict the nonlinear evolution of the bulk modulus, being most of then introduced in the small-strain framework or the Equivalent Medium Theory (EMT) \cite{Goddard1990,Walton1987,Zaccone2011_Approximate,LaRagione2012}. 
That approach adopts an analogue model considering a set of springs joining the center of mass of bodies in contact and whose deformation represents the relative approaching of their centers as they deform. 
Using this type of approach, it is then possible to integrate the set of springs' deformation and deduce a stress-strain relation for the whole system; thus, a bulk equation can also be deduced. 
Nonetheless, the definition of the spring behavior is, in that vein, at the origin of the macroscopic strains and bulk evolution. 
Similarly, smooth discrete-element methods can consider the force-overlapping relations to be the spring characteristics for an equivalent medium analogy. 
If we use a similar strategy, we can first consider $\ell$ to be the average spring length and $\varepsilon_{\ell} = \ln(\ell /d)$ to be the average strain, with $d$ being the average particle diameter (our numerical experiments also show that $\varepsilon_v = 4 \varepsilon_{\ell}$). 
Second, the homogeneous field of springs allows us to suppose that the contact level pressure is $P_{\ell} = E \varepsilon_{\ell}$. 
The granular stress tensor can be written as $\sigma_{ij} = n_c \langle f_i \ell_j \rangle$, with the contact number density being $n_c = N_c/V$, $N_c$ is the number of contacts, $V$ is the sample volume, and $\langle \dots \rangle$ the average of contact forces $\bm{f}$ and branch vectors $\bm{\ell}$ (i.e., the inter-center vector of particles in contact).
We can then deduce a micromechanical definition of pressure as 
\begin{equation}\label{eq:P}
P = \frac{Z\phi}{\pi} \sigma_{\ell},
\end{equation}
with the coordination number $Z=2N_c/N_p$ and $\sigma_\ell = \langle f\ell\rangle/d^2$. 
Note that to deduce the expression above, we consider $P=(\sigma_1 + \sigma_2)/2$, where $\sigma_1$ and $\sigma_2$ are the principal stresses of tensor $\sigma_{ij}$.
If we consider that the coordination number evolves as a power-law of the solid fraction (as consistently shown in previous studies as \cite{Vu2019,Nezamabadi2019,Andreotti2013}, and also verified in our simulations), in the form $(Z-Z_0) = k (\phi - \phi_0)^{\alpha}$, with $\alpha = 0.5$, $Z_0$ the coordination number in the reference state, and $k$ a proportionality parameter easily deduced knowing that, when $\phi$ tends to unity, the particle structure tends to a hexagonal-like arrangement and $Z \rightarrow 6$. 
So, $k\simeq 5$. 
Finally, considering that $\sigma_\ell = P_\ell$, we can deduce a microscopic definition of the bulk modulus upon the derivative of Eq. (\ref{eq:P}) as 
\begin{equation}\label{eq:k1}
K/E = \frac{Z \phi}{4\pi} \left( \frac{5}{2} - \frac{\phi_0}{\phi}\right) - \frac{Z_0 \phi}{8\pi}. 
\end{equation}

This equation is displayed in Fig. \ref{fig:k_and_eqs} with a solid blue line. 
We can observe that the predictions with such an expression can be considered relatively good for the first part of the compression, and up to $\phi \simeq 0.85$. 
Beyond that value, the bulk modulus starts to increase more rapidly and is expected to diverge as the solid fraction tends to unity (i.e., the sample ends up behaving as an incompressible solid). 
The equivalent medium composed of springs cannot capture the diverging behavior of the bulk. 
It is necessary then to track the evolution of $K$ for the multi-particle system using an alternative approach. 

\section{Single-particle scale }\label{Local}
\begin{figure}
    \centering
    \includegraphics[width=0.9\linewidth]{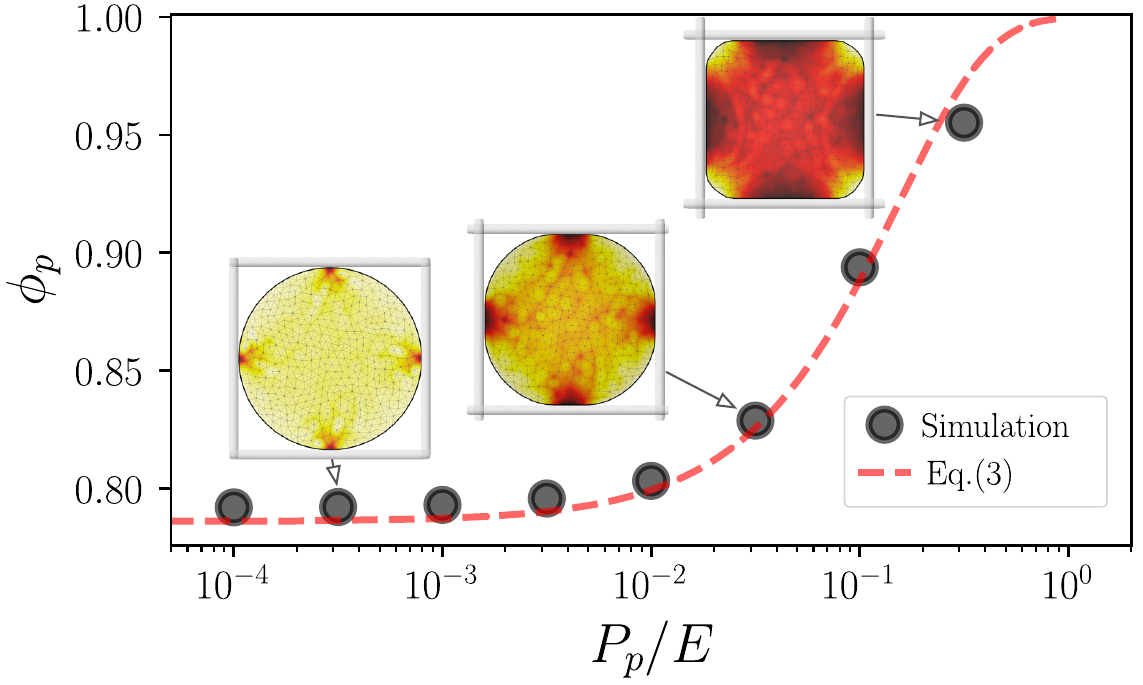}
    \caption{Compaction behavior of a single particle under isotropic compression. The dashed red line shows the analytical expression in Eq. (\ref{eq:Pp}) nicely fitting the results of our simulations. Screenshots show the deformed state of the particles inside the squared box.} 
    \label{fig:p_phi_part}
\end{figure}

Let us consider the system composed of a single soft particle inside a square box and following the preset boundary conditions as undertaken with the assembly. 
Figure \ref{fig:p_phi_part} presents the evolution of the corresponding solid fraction ($\phi_p$) as a function of the applied pressure $P_p$. 
Note that the volumetric strain for this case is akin to the multi-particle system but denoted $\varepsilon_{v,p}$. 
The compression behavior of the single-particle test can be considered analogous to the collapse of a cavity inside a circular body employing a homogeneous external pressure \cite{Carroll1984}. 
For that case, elastic solutions allow us to write the relation $P_p-\phi_p$ as 
\begin{equation}\label{eq:Pp}
P_p/E = -b \ln\left(\frac{\phi_{p,max}-\phi_p}{\phi_{p,max}-\phi_{p,0}}\right),
\end{equation}
with $\phi_{p,max}$ the solid fraction as $P_p \rightarrow \infty$, and $\phi_{p,0} = \pi/4$ the solid fraction at the reference state (i.e., as $P_p \rightarrow 0$). 
Finally, parameter $b$ is found out to be $\simeq 0.12$ after fitting Eq. (\ref{eq:Pp}) to our data. 
We can observe that this expression fits very well the behavior of the single-particle compression as it is plotted with a red dashed line in the same Fig. \ref{fig:p_phi_part}.

We can then easily deduce an analytical bulk modulus equation by using the derivative of the last expression, yielding to $K_p/E = (b\phi_p)/(\phi_{p,max} - \phi_p)$. 
In sum, at single-particle scale, we can comprehensibly identify analytical equations for both the compression and bulk evolution. 

\section{Conciliating scales}\label{Model}
Let us consider again Eq. (\ref{eq:k1}) on the macroscopic bulk modulus. 
In the limit of $\phi\rightarrow\phi_0$, and in agreement with many previous works \cite{Goddard1990,Walton1987,Zaccone2011_Approximate,LaRagione2012}, 
that equation shows that $K \propto Z\phi E$.
This means that, for small deformations, the bulk modulus scales via the Young modulus of a single particle and the structural parameters $Z$ and $\phi$ of an assembly of particles. 
Naively, we may wonder whether the single-particle bulk modulus $K_p$ is more representative than the Young modulus $E$ for such scaling. 
In other words, is there a more general scaling between the bulk modulus of the multi-particle and the single-particle system?
In fact, by comparing these two systems at equivalent deformation (i.e., $\varepsilon_v = \varepsilon_{v,p}$), our numerical simulations reveal that
\begin{equation}\label{eq:k0}
K/E \simeq \frac{Z\phi}{2\pi} K_p. 
\end{equation}

Equation (\ref{eq:k0}) points out that the single-particle configuration can indeed be considered the smallest representative scale in our multi-particle system.
Now, by replacing $K_p$ by its analytical form identified before and by mapping $\phi$ with $\phi_p$ (since $\varepsilon_v = \varepsilon_{v,p}$),
we get a general micro-mechanically based constitutive equation for the bulk modulus evolution beyond the jamming as 
\begin{equation}\label{eq:k2}
K/E \simeq \frac{b \phi^2}{2\pi(\phi_{max} - \phi)} \{Z_0 + k(\phi - \phi_0)^{\alpha}\}. 
\end{equation}

Equation (\ref{eq:k2}) is shown in Fig. \ref{fig:k_and_eqs} showing an excellent agreement with the measured $K$ in our tests. 
In this case, the equation follows the nonlinear increase of $K$ for large deformations and its divergence. 

\section{Effect of the coefficient of friction}\label{Friction}
The inter-particle coefficient of friction is a mechanical constraint rapidly limiting the particle reorganization and, thus, their capacity of filling voids. 
We reproduce a series of tests with the same samples as for the frictionless case, but setting this time the interparticle coefficient of friction to $\mu_p = \{0.2, 0.4, 0.6, 0.8\}$. 
We tracked the evolution of pressure and deformation, and finally, we could compute the evolution of $K$ for these cases.

\begin{figure}
    \centering
    \includegraphics[width=0.9\linewidth]{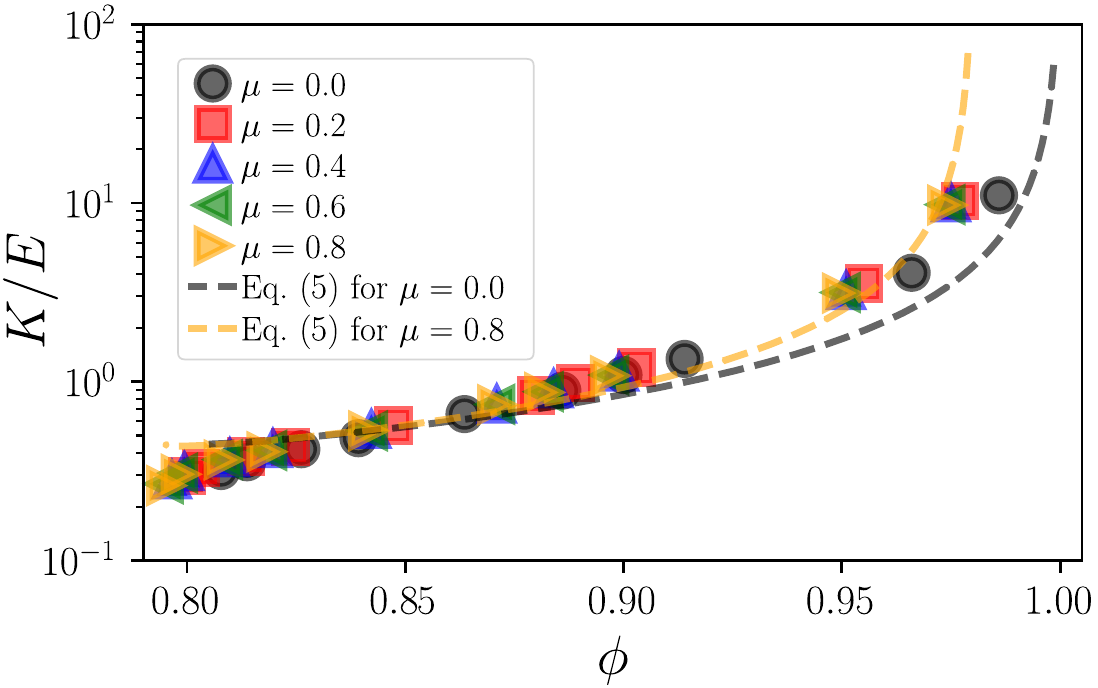}
    \caption{Evolution of the macroscopic bulk modulus as a function of the solid fraction for different values of interparticle coefficient of friction $\mu$. The dashed lines are drawn using Eq. (\ref{eq:k2}) fitting cases $\mu=0.0$ (black) and $\mu=0.8$ (orange).} 
    \label{fig:k_phi}
\end{figure}

Figure \ref{fig:k_phi} gathers the results for the evolution of $K$ as a function of the assembly's density for the different values of $\mu_p$. 
We can observe that for small increments of solid fraction after jamming, the evolution of $K$ is indistinguishable for the different coefficients of friction. 
Nonetheless, for higher values of solid fraction (i.e., larger deformations), the data gathered from the simulations slightly shifts towards the left as $\mu_p$ increases. 
This means that the additional constraint added by friction limits the deformability of the assembly. 
Many authors have shown that friction modifies the jammed state of rigid grains assemblies and the maximal solid fraction a soft assembly of grains can reach \cite{VanHecke2010,Silbert2006}. 
Consequently, the reference state, as $P \rightarrow 0$, and the limit state, as $P \rightarrow \infty$, are modified with respect to the frictionless case. 
These facts indeed affect the proposed model in Eq. (\ref{eq:k2}). 
While for the frictionless case $Z_0 \simeq 3.9$, $\phi_0 \simeq 0.805$, and our fitting procedure resulted in $\phi_{max} \simeq 0.998$, the case with $\mu=0.8$ shows that $Z_0 \simeq 3.6$, $\phi_0 \simeq 0.795$, and fitting of the data ends up estimating $\phi_{max} \simeq 0.978$. 
Using these parameters, we plot in the same figure the predictions of bulk evolution using our model with dashed lines. 
We can observe that the fitting is still excellent, and we can follow the macroscopic evolution of the bulk modulus up to densities very close to unity. 

\section{Conclusions}\label{Conclu}
We studied the bulk modulus behavior of two-dimensional particle grain assemblies undergoing isotropic compression using a coupled discrete-finite element simulation platform. 
By finely describing the assemblies' compaction behavior, we presented the bulk modulus evolution as a function of the solid fraction. 

We first introduced a model approach for the bulk modulus evolution standing on the ideas of the equivalent medium approach and considering the soft particle assembly as a series of interconnected springs governed by the particles elastic modulus $E$. 
We observed that such an approach leads to a formulation of the bulk modulus evolution that is acceptable only for relatively small strains and does not reproduce the divergence of $K$ as the density approaches unity. 
To overcome this challenge, we first studied the compression of a single particle under isotropic compression in a box. 
We observed that the compression behavior and bulk evolution for such a case could be analytically obtained based on elastic solutions of a circular geometry under compression. 
Then, standing on the definition of the granular stress tensor and supposing that the most elemental scale of the assemblies is the single-particle, we conciliated the single and multi-particle scales and deduced a model equation for $K$. 
Our model turned out to reproduce the evolution of the bulk modulus very well, highlighting its strong non-linearities and asymptotes. 
We finally showed that it is straightforward to introduce friction between particles into our model and then consider the impact this mechanical constraint has on limit configurations the system can reach. 

The reader can also deduce that our modeling approach lets us describe the compaction evolution the writing of a relationship $P$ vs. $\phi$. 
This constitutive description can be found in Ref. \cite{Cantor2020} by the same authors of this work, or in Ref. \cite{Cardenas2020} taking into account mixtures of rigid and deformable particles. 
Nonetheless, to enrich the constitutive modeling of granular materials and structures upon discrete systems, it is still necessary to explore the shear behavior of soft particle media. 
It is also important to extend this work to varied shape and size particles, as well as to the 3D case. 

\bibliography{biblio}

\end{document}